# Characterizing AI-Assisted Bot Traffic in Darknet Data: Implications for ICS and IIoT Security

Alex Carbajal*, Caleb Faultersack*, Jonahtan Vasquez*, Shereen Ismail†, and Asma Jodeiri Akbarfam*
*School of Engineering and Applied Sciences, Washington State University Tri-Cities, Richland, WA 99354, USA
†Merit Network, Inc., University of Michigan, Ann Arbor, MI 48108, USA

***Abstract*—The rise of automated scanning tools and AI-assisted reconnaissance agents has significantly altered internet background traffic patterns, threatening the baseline assumptions underlying intrusion detection systems (IDS) deployed in critical infrastructure networks. This paper characterizes the evolution of automated bot traffic by analyzing a longitudinal dataset of 192 million passive darknet packets captured across 2021 and 2025 from the Merit ORION Network Telescope. A modular analysis pipeline was developed to compute metrics including average packet rate, global Shannon entropy, inter-arrival time (IAT) burstiness, geographic attribution, and destination port targeting across key industrial protocols. Results reveal a highly distributed yet focused reconnaissance landscape, with traffic targeting ICS-relevant ports nearly doubling from 0.82% to 1.51% over the four-year period. Furthermore, burstiness analysis exposes intentional micro-pacing behaviors (1ms to 100ms delays) that allow modern botnets to artificially smooth their overall volume. Our simulated anomaly-based IDS demonstrates that these evasion techniques enable 97.47% of modern bot traffic to bypass standard volumetric thresholds undetected. Compensatory sensitivity tuning triggers a 68.10% false-positive rate, highlighting fundamental visibility and alerting gaps in operational technology (OT) environments.**

## I. Introduction

The landscape of unsolicited internet traffic has undergone a significant transformation over the past decade. What was once dominated by opportunistic worms, misconfigured devices, and indiscriminate port scanners has evolved into a more sophisticated ecosystem of automated reconnaissance tools, AI-assisted vulnerability scanners, and intelligent botnet campaigns [1], [2] with recent studies showing that this traffic exhibits structured and centralized behavior rather than random patterns [3]. Recent industry reporting indicates that automated bot traffic now constitutes over 51% of all web traffic, with AI-driven bots accounting for a growing share of that activity, a trend with direct implications for network security monitoring [4]. These changes are not merely academic; they have concrete consequences for the security of systems that underpin modern digital systems.

Critical infrastructure networks, including SCADA systems, distributed control systems (DCS), and Industrial Internet of Things (IIoT) deployments, have become increasingly interconnected with broader IP networks [5]. This connectivity, while operationally beneficial, exposes industrial control system (ICS) components to the same background scanning activity that pervades the general internet. Unlike traditional IT environments, OT networks prioritize availability and physical safety above all else, making the consequences of a successful intrusion far more severe than a typical data breach [5]. A compromised water treatment facility, power substation, or manufacturing process can result in physical damage, service disruption, or even loss of life [6].

Intrusion detection systems (IDS) deployed in operational technology (OT) environments rely on carefully calibrated baselines derived from the assumption that background internet traffic behaves in a statistically predictable manner. Anomaly-based IDS approaches flag deviations from this established norm as potential threats. However, if the nature of background traffic itself has shifted, becoming more structured, more targeted, or more temporally varied due to automation and large-scale scanning, then these baseline assumptions may no longer hold [7], [8]. The result is a degradation in IDS reliability: increased false positive rates as benign automated traffic triggers alerts, and reduced visibility as sophisticated bot activity actively exploits these assumptions, smoothly pacing traffic to blend into the noise [9], [10]. These findings align with broader trends in cyber–physical security research showing that compromised or manipulated telemetry can significantly degrade operational visibility and situational awareness in critical infrastructure environments [11]. Similar trends in infrastructure security have also shown that legacy or misconfigured deployments continue to expose operational environments to disruption-oriented attacks [12].

Despite a growing body of research on internet background radiation and darknet traffic, the specific impact of automated and AI-assisted bot traffic on OT-adjacent network behavior remains underexplored. Prior work has characterized large-scale darknet traffic using metrics such as Shannon entropy, inter-arrival distributions, and scanning strategy [7], [13], and recent studies have tracked temporal changes in scanning behavior using data from large network telescopes [9], [14]. However, few studies have explicitly examined how these metrics have shifted longitudinally in response to the proliferation of automated tooling, nor have they connected these shifts directly to the threat landscape of ICS and SCADA environments [6], [13].

This paper addresses that gap. Using passive network telescope data from the Merit ORION telescope [14], [15], we analyze a longitudinal, distributed sample of 192 million packets spanning 2021 and 2025. We characterize behavioral shifts in data volume, global Shannon entropy, inter-arrival time (IAT) burstiness, geographic origin, and the concentration of activity on known industrial protocol ports such as Modbus (TCP/502), DNP3 (TCP/20000), EtherNet/IP (TCP/44818), and IEC 60870-5-104 (TCP/2404). While previous studies have characterized darknet traffic broadly, our unique contribution is isolating the specific evolution of ICS-targeted reconnaissance, demonstrating that the fraction of traffic targeting these critical ports has nearly doubled from 0.82% to 1.51%. Furthermore, by evaluating this empirical traffic against a simulated IDS, we prove that the artificial pacing of modern bots allows over 97% of malicious traffic to successfully evade standard volumetric baselines, fundamentally challenging current anomaly-based detection paradigms.

The remainder of this paper is organized as follows: Section II reviews background and related work on darknet traffic analysis and ICS security monitoring. Section III describes the datasets and methodology. Section IV presents our system architecture and implementation. Section V discusses our results. Section VI explains our plan for future work, and Section VII concludes the paper with a summary of current progress and future direction.

## II. Background / Context

Darknet or network telescope measurements provide a powerful lens into unsolicited internet traffic by monitoring packets destined for routable but unused IP address space. Because no legitimate services operate on these addresses, nearly all observed traffic corresponds to scanning, misconfiguration, exploitation attempts, or attack backscatter rather than normal user activity [4]. Prior work has used large network telescopes such as CAIDA's UCSD network telescope and Merit's ORION telescope to study internet background radiation, revealing long-term trends in scanning intensity, protocol usage, and the emergence of large-scale botnet activity [2], [7], [8], [9], [10], [13], [16].

At the same time, critical infrastructure networks built around ICS and OT have become increasingly interconnected with enterprise IT and the public internet [5]. Systems such as SCADA, DCS, and IIoT deployments expose services that implement industrial protocols like Modbus, DNP3, EtherNet/IP, IEC 104, and S7/ISO-TSAP over TCP or UDP, often through perimeter firewalls or DMZs. Unlike traditional IT environments that emphasize confidentiality and integrity, these OT environments prioritize availability and physical safety, meaning that successful attacks can cause process disruption, equipment damage, or physical harm rather than only data loss.

IDS deployed in OT environments typically rely on a combination of signature-based and anomaly-based techniques [6]. Signature-based IDS identify known malicious patterns, whereas anomaly-based systems establish statistical baselines for "normal" traffic and flag deviations as potential threats. Many of these baselines implicitly assume that background internet traffic is relatively random and stationary over time, with noise characterized by broad, uniform scanning rather than structured or adaptive behavior. If background traffic grows more coordinated, targeted, or bursty, these assumptions may erode, increasing false positive rates and allowing sophisticated automated activity to blend into the noise [17].

Recent industry and academic work [4], [6], indicates that automated and AI-assisted bot traffic now constitutes a substantial fraction of overall internet activity and plays an increasingly important role in background scanning behavior. Modern bots can dynamically adapt their scanning strategies, prioritize high-value ports, and mimic benign traffic patterns, complicating traffic modeling and IDS tuning [3]. Several studies [2], [7], have analyzed darknet traffic using metrics such as Shannon entropy, inter-arrival distributions, and spatial-temporal correlation to characterize scanning campaigns, detect structural changes, and distinguish between random and organized activity.

For ICS/OT security, these developments are especially concerning because adversaries frequently probe industrial protocol ports exposed, intentionally or accidentally, to the public internet [6]. Traffic directed at ports like 502 (Modbus), 20000 (DNP3), 44818/2222 (EtherNet/IP), 2404 (IEC 104), 102 (S7/ISO-TSAP), 47808 (BACnet), and 4840 (OPC UA) can reveal the presence and configuration of industrial devices, enabling targeted attacks. Understanding how automated bot traffic shapes darknet observations on these ports is therefore critical for assessing exposure and for designing IDS strategies that remain effective as background traffic evolves.

Against this backdrop, the present work situates itself at the intersection of darknet measurement, AI-driven bot behavior, and ICS/OT security. By characterizing traffic observed by the Merit ORION telescope in terms of packet rate, entropy, burstiness, port targeting, and geographic origin, with an emphasis on ICS-relevant ports, the study aims to provide actionable context for operators and researchers seeking to calibrate IDS baselines and interpret alerts in environments increasingly saturated with automated scanning.

## III. Methodology / Approach

This study combines passive darknet measurement, feature extraction, and statistical analysis to characterize automated bot traffic targeting ICS/OT environments. All experiments utilize PCAP data collected from the Merit ORION Network Telescope together with temporally aligned MaxMind GeoLite2 databases for geographic attribution.

The dataset consists of 192 million passive darknet packets sampled across 2021 (baseline) and 2025 (test dataset). To provide seasonally distributed coverage, traffic was extracted from the 15th day of every month at four UTC intervals (00:00, 06:00, 12:00, and 18:00). Each interval processed 2 million packets, yielding 96 million packets per year. Packet records included timestamps, source and destination IP addresses, transport-layer ports, and protocol identifiers.

To reduce inaccuracies caused by IP reallocations over the four-year period, separate GeoLite2-Country `.mmdb` databases corresponding to each collection year were used. Data preprocessing and analysis were implemented in Python using

`dpkt` for packet parsing and `pandas`/`numpy` for statistical processing.

The analysis pipeline first parsed PCAP files, removed malformed or non-IP packets, and generated comparative traffic statistics including packet rate, bandwidth, and protocol activity. ICS/OT analysis focused on industrial ports associated with Modbus (502), DNP3 (20000), EtherNet/IP (44818/2222), IEC 104 (2404), and S7/ISO-TSAP (102). Scanning behavior was evaluated using dynamically scaled destination IP gap analysis to distinguish sequential sweeps from randomized probing.

To characterize structural traffic changes, global Shannon entropy was computed for both source IPs and destination ports. Burstiness analysis utilized Inter-Arrival Time (IAT) distributions accumulated into logarithmic bins spanning $10^{-3}$ to $10^{3}$ milliseconds, enabling efficient detection of artificial packet pacing behavior across large-scale traffic volumes.

Geographic attribution was used to measure cross-year changes in source-country activity and identify regional concentration shifts in automated scanning infrastructure. Additional visualization modules generated comparative ICS port activity plots and traffic delta distributions.

To evaluate operational implications, 1-second packet-rate time series were extracted and tested against a simulated volumetric anomaly-based IDS. Detection thresholds were derived from the 2021 baseline using a 99.7% confidence interval (Mean + 3 Standard Deviations). The 2025 dataset was then evaluated against this baseline to measure botnet evasion capability and false-positive behavior under high-sensitivity tuning.

All analysis scripts and generated outputs are publicly available to support reproducibility.[1]

## IV. System Design / Implementation

This section describes our modular, reproducible analysis pipeline for characterizing unsolicited darknet traffic from the Merit ORION Network Telescope (Fig. 1).

The primary input consists of a longitudinal dataset of 192 million packets systematically captured across 2021 and 2025. Monitoring approximately 500,000 unallocated IPv4 addresses [14], this telescope captures pure background radiation (e.g., port scans, exploit attempts, backscatter). Raw PCAPs are ingested utilizing the `dpkt` library for high-performance packet parsing [18]. Malformed or non-IP packets are discarded, and timestamps are strictly normalized to UTC to ensure cross-year consistency.

From the valid IP packets, the pipeline extracts the following key features: source IP (for geographic attribution), destination port (for protocol identification), transport protocol, and timestamp. These features drive three core statistical evaluations:

1) **Global Shannon Entropy**: Compares actual source IP and destination port entropy distributions to measure traffic diversity. High source IP entropy combined with low destination port entropy indicates a highly distributed botnet targeting specific services.
2) **Burstiness (IAT Distributions)**: Evaluates the Inter-Arrival Times (IAT) between consecutive packets. To process 100-million-packet intervals without memory degradation, the system utilizes a pre-binned accumulation strategy, dropping IATs directly into logarithmic bins to isolate the 1ms to 100ms artificial packet pacing characteristic of evasion.
3) **Scanning Strategy Gaps**: Calculates the numerical distance between targeted IPs using dynamically scaling thresholds, distinguishing highly concentrated sequential sweeps from randomized, opportunistic probing.

To isolate threats to critical infrastructure, traffic is flagged against a lookup table of 17 OT-relevant ports, including Modbus (TCP/502), DNP3 (TCP/20000), EtherNet/IP (TCP/44818 and 2222), S7/ISO-TSAP (TCP/102), OPC UA (TCP/4840), BACnet (UDP/47808), and SNMP [19]. Simultaneously, source IPs are mapped to their geographic origins using temporally aligned MaxMind GeoLite2-Country databases [20] to prevent historical IP reassignment errors.

The pipeline is fully parameterized via a central bash orchestrator (`run_yearly_analysis.sh`) for extensibility. Independent modules process the extracted data and output structured CSVs and high-resolution visualizations for the final analysis, including: a core metrics summary table tracking the dominant ICS protocols, ICS scanning pattern charts, entropy and burstiness distributions, a geographic delta shift table, a connected dumbbell plot isolating top ICS port volume shifts, and an anomaly-based IDS simulation demonstrating volumetric threshold evasion.

## V. Results

The analysis pipeline has been fully implemented and applied to a longitudinal dataset consisting of 192 million packets captured from the Merit ORION Network Telescope. This dataset, sampled systematically across 2021 and 2025, provides a comprehensive view of contemporary unsolicited internet traffic and enables robust characterization of automated scanning behavior over time.

### A. Dataset Characterization:

As summarized in Table I, comparative exploration of the PCAP data shows a clear shift in scanning focus. While broad opportunistic scanning across well-known IT services continues, the proportion of traffic directed toward industrial control system (ICS) ports nearly doubled from 0.8156% in the 2021 baseline to 1.5064% in the 2025 bot traffic.

Despite the continued presence of general-purpose scanning, this measurable surge directed toward ICS ports is highly significant. To put this into context, if scanning were purely random across all 65,535 available ports, the 17 monitored ICS ports would be expected to receive only about 0.026% of the traffic. The 1.51% observation in 2025 represents a massive, multi-fold increase over the random baseline, strongly confirming that operational technology relevant services are being actively and specifically probed rather than merely caught in indiscriminate sweeps.

[1] https://github.com/alex-sir/bot-traffic

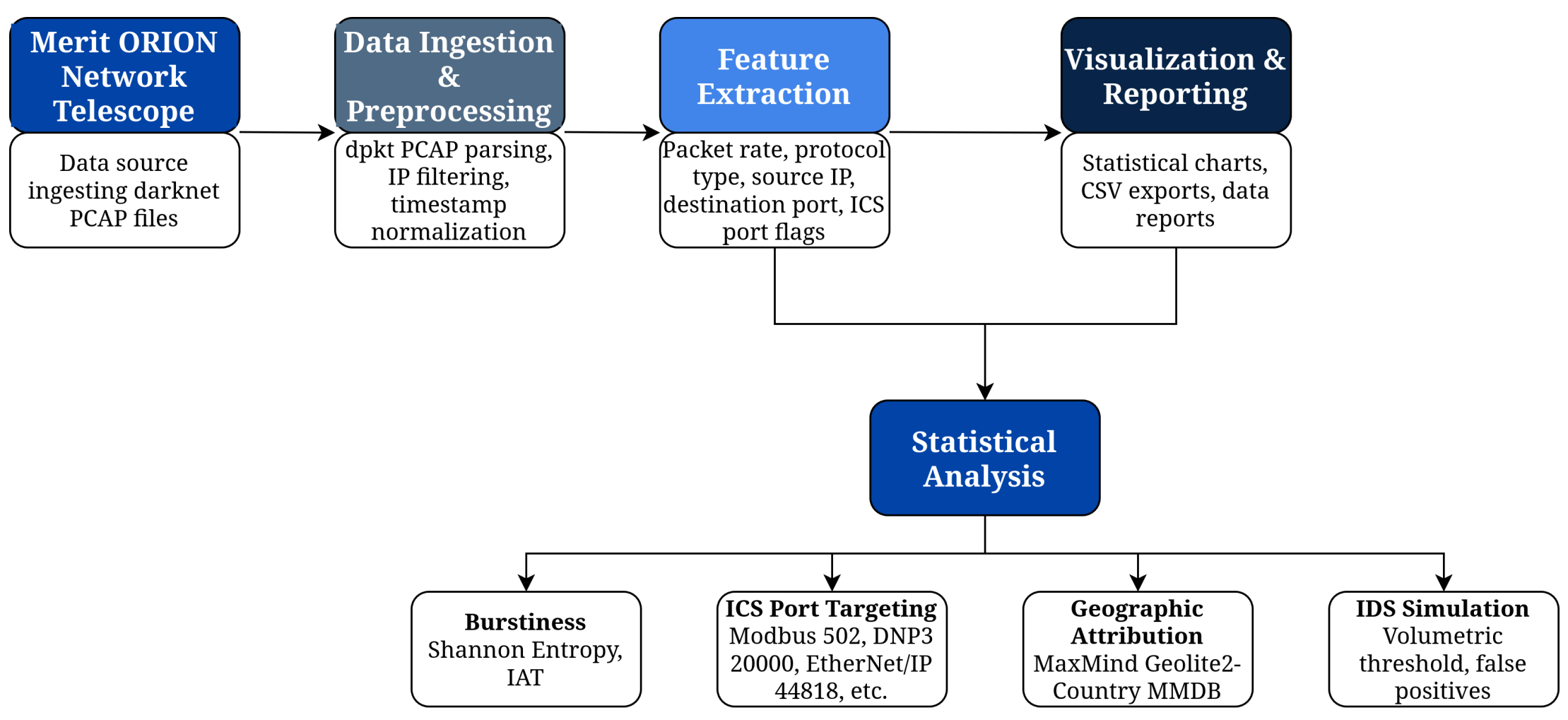

Fig. 1. System Architecture for Darknet Traffic Analysis Pipeline

TABLE I
CROSS-YEAR PCAP OVERVIEW STATISTICS

| Metric | 2021 Baseline | 2025 Bot Traffic |
|---|---|---|
| **Files Analyzed** | 48 | 48 |
| **Initial Start (UTC)** | 2021-01-15 00:00:00 | 2025-01-15 00:00:00 |
| **Active Duration** | 2546.4 sec | 2186.8 sec |
| **Total Packets** | 96,000,000 | 96,000,000 |
| **Total Volume** | 6546.54 MB | 7055.00 MB |
| **Avg Packet Rate** | 37700.4 pkts/s | 43900.8 pkts/s |
| **Avg Bandwidth** | 21.566 Mbps | 27.064 Mbps |
| **Dominant ICS Protocol** | SNMP (ICS mgmt) | EtherNet/IP (alt) |
| **ICS Traffic** | 0.8156% (782,953) | 1.5064% (1,446,160) |
| **Non-ICS Traffic** | 99.1844% (95,217,047) | 98.4936% (94,553,840) |
| **Unique Src IPs** | 1,263,320 | 560,907 |
| **Unique Dst IPs** | 475,136 | 475,648 |
| **Unique Dst Ports** | 65,536 | 65,536 |

### B. ICS Port Targeting Behavior:

Analysis of ICS specific ports reveals distinct scanning patterns, illustrated in Fig. 2. Ports associated with protocols such as EtherNet/IP, BACnet, and S7/ISO-TSAP exhibit the highest levels of activity among ICS targets. While overall volume remains lower than traditional IT ports, the presence of consistent probing suggests persistent reconnaissance of industrial systems.

Scanning strategy analysis indicates a mix of behaviors. For some ports, low average address gaps suggest sequential scanning, consistent with systematic sweeping by automated tools. In contrast, higher gap values indicate more randomized probing, likely associated with distributed reconnaissance or large scale data collection efforts. These findings suggest that automated scanning campaigns are not purely random, but instead combine structured and opportunistic strategies depending on the target protocol.

### C. Entropy and Traffic Dynamics:

Statistical analysis of traffic behavior reveals a structural shift indicative of modern, distributed botnets. The global Shannon entropy for source IPs increased in the 2025 dataset, indicating that scanning activity is originating from a wider, more diverse swarm of compromised hosts. Conversely, destination port entropy decreased, reflecting a shift away from 'shotgun' internet scanning toward targeting of specific services (see Fig. 3).

Burstiness analysis based on Inter-Arrival Times (IAT) further exposes the evasion tactics of these automated agents. Rather than blasting packets at maximum capacity, the 2025 traffic exhibits highly intentional micro-pacing, visualized as a distinct frequency spike in the 1ms to 100ms IAT window (see Fig. 4). These artificial delays flatten the overall traffic volume curve, challenging the assumption of stationary background traffic often used in anomaly based intrusion detection systems.

### D. Temporal Port Activity:

Beyond aggregate volume, visualizing the net shifts in scan intensity reveals precisely how industrial protocol targeting has evolved. As illustrated in the port activity matrix (Fig. 5), probes against specific operational technology ports have surged dramatically over the four-year gap. The most significant absolute increases occurred on EtherNet/IP (alt) on port 2222, which saw an influx of over 339,000 packets, alongside targeted increases against S7/ISO-TSAP and ProConOS. This confirms that industrial reconnaissance tends to occur in concentrated, protocol-specific bursts rather than continuous, uniform sweeps.

### E. Geographic Distribution:

Geographic attribution of source packets highlights a radical shift in the regional origin of scanning activity between 2021 and 2025. Table II shows that while historical sources of darknet traffic like Russia (-88.3%) and Iran (-97.4%) saw massive declines, new infrastructures emerged rapidly. The concentration of traffic shifted heavily toward the United States (+77.1%) and Europe, with explosive percentage increases originating from cloud infrastructure and potentially compromised devices in Bulgaria (+1166.5%) and Romania (+1001.8%). An anomaly was also observed in the Seychelles (+2058.1%). This concentration suggests that scanning activity is not uniformly distributed across the internet, but instead relies on specific regional clusters.

TABLE II
TOP 15 SOURCE COUNTRIES BY TOTAL PACKET VOLUME

| Country | 2021 Baseline Pkts | 2025 Bot Traffic Pkts | Δ (%) |
|---|---|---|---|
| **United States** | 18,145,977 | 32,131,467 | +77.1% |
| **Russia** | 27,718,463 | 3,229,316 | -88.3% |
| **United Kingdom** | 12,944,336 | 5,253,416 | -59.4% |
| **China** | 9,622,027 | 4,649,432 | -51.7% |
| **Netherlands** | 3,362,539 | 9,679,097 | +187.9% |
| **Bulgaria** | 739,657 | 9,367,441 | +1166.5% |
| **Romania** | 552,416 | 6,086,551 | +1001.8% |
| **Iran** | 6,429,307 | 169,095 | -97.4% |
| **Germany** | 1,857,255 | 4,673,309 | +151.6% |
| **Canada** | 512,094 | 2,350,185 | +358.9% |
| **France** | 596,824 | 2,045,527 | +242.7% |
| **Ukraine** | 723,697 | 1,897,788 | +162.2% |
| **Singapore** | 1,132,854 | 1,113,145 | -1.7% |
| **Hong Kong** | 709,054 | 1,491,215 | +110.3% |
| **Seychelles** | 72,760 | 1,570,242 | +2058.1% |

### F. Anomaly-Based IDS Simulation:

To practically measure the impact of these behavioral shifts on critical infrastructure defenses, the 2025 bot traffic was evaluated against a simulated volumetric anomaly-based IDS trained on the 2021 baseline. Using a strict 99.7% confidence interval (Mean + 3 Standard Deviations), the standard baseline threshold was calculated at 57,102 packets per second.

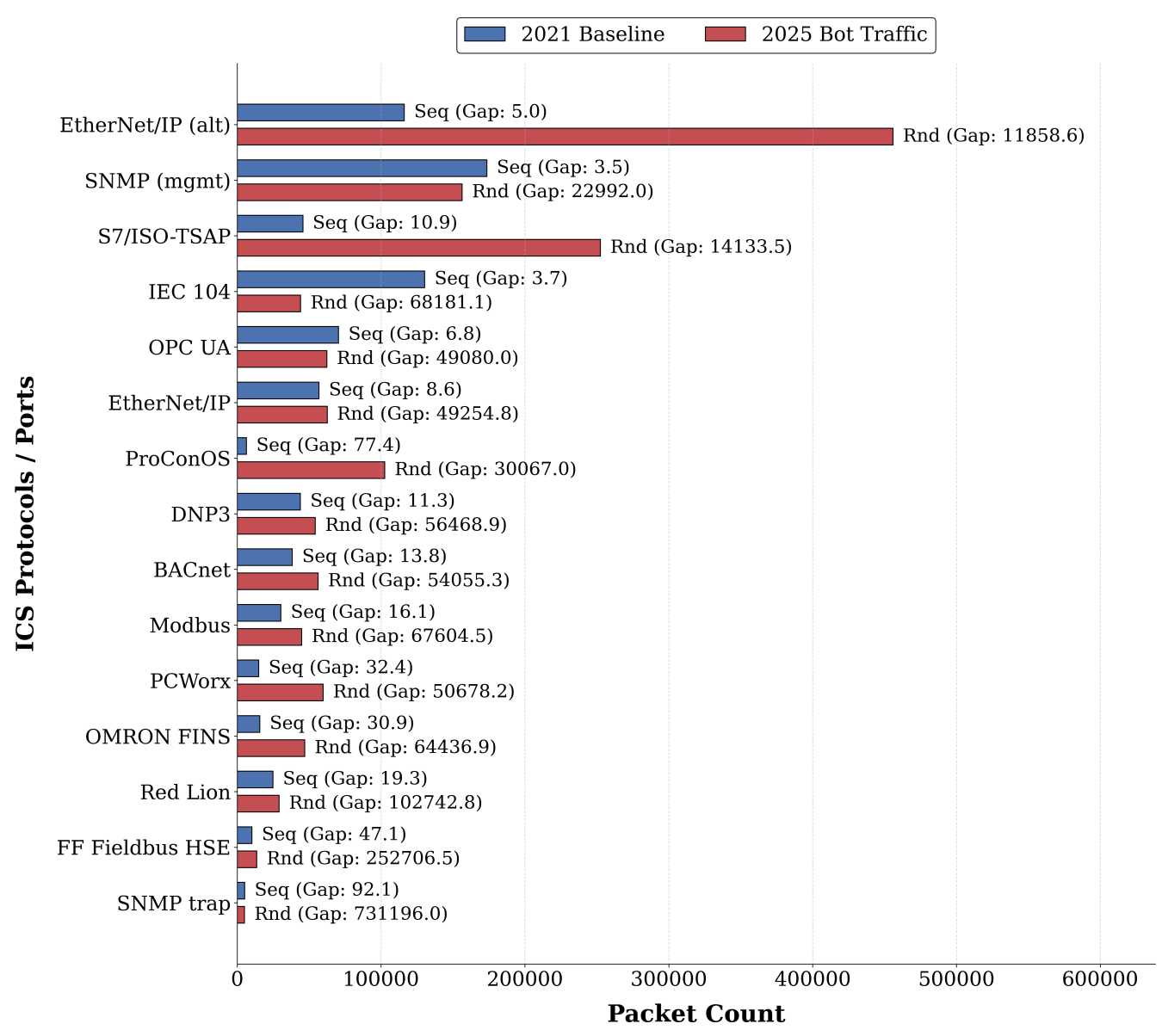

Fig. 2. ICS Port Targeting Volume and Identified Scanning Patterns

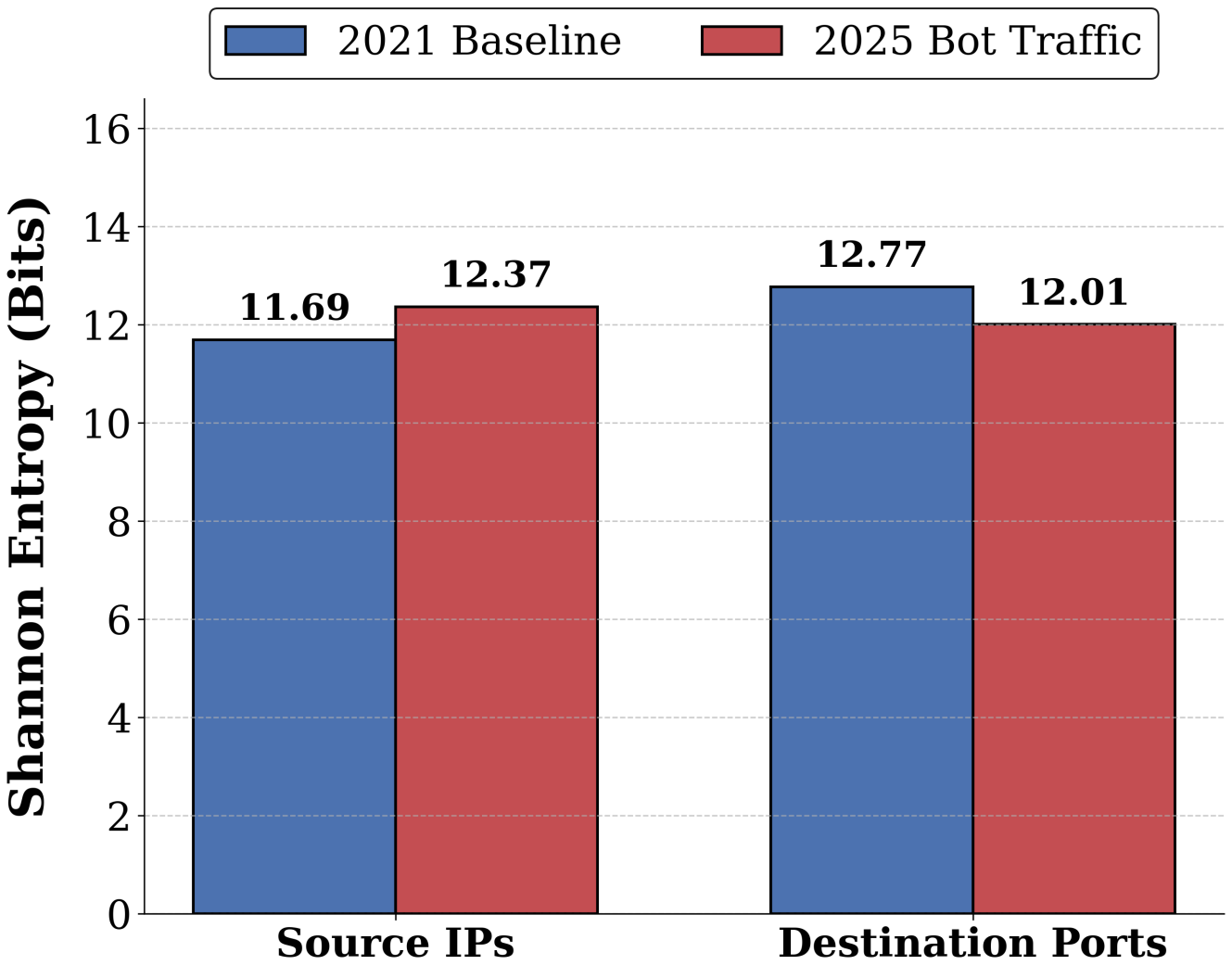

Fig. 3. Cross-year Shannon Entropy Comparison

As shown in Fig. 6, because of the distributed nature and deliberate 1ms to 100ms pacing of the AI-assisted botnet, the malicious traffic successfully hid within the historical variance of the network. The standard threshold yielded a 97.47% evasion rate, detecting only 2.53% of the bot traffic.

Attempts to rectify this by implementing high-sensitivity tuning proved equally unviable. Lowering the trigger threshold to 34,928 packets per second successfully achieved a 90% detection rate for the botnet; however, it simultaneously flagged 68.10% of the normal baseline traffic as anomalous. This explicitly proves that static volumetric thresholding is fundamentally incapable of separating intelligently paced botnet scans from benign background variance without inducing severe alert fatigue.

### G. Emerging Patterns and Implications:

Across all analyses, a consistent theme emerges: modern background internet traffic is no longer defined by noisy, opportunistic worms, but by stealthy, highly structured botnets. Indicators of this structure include:

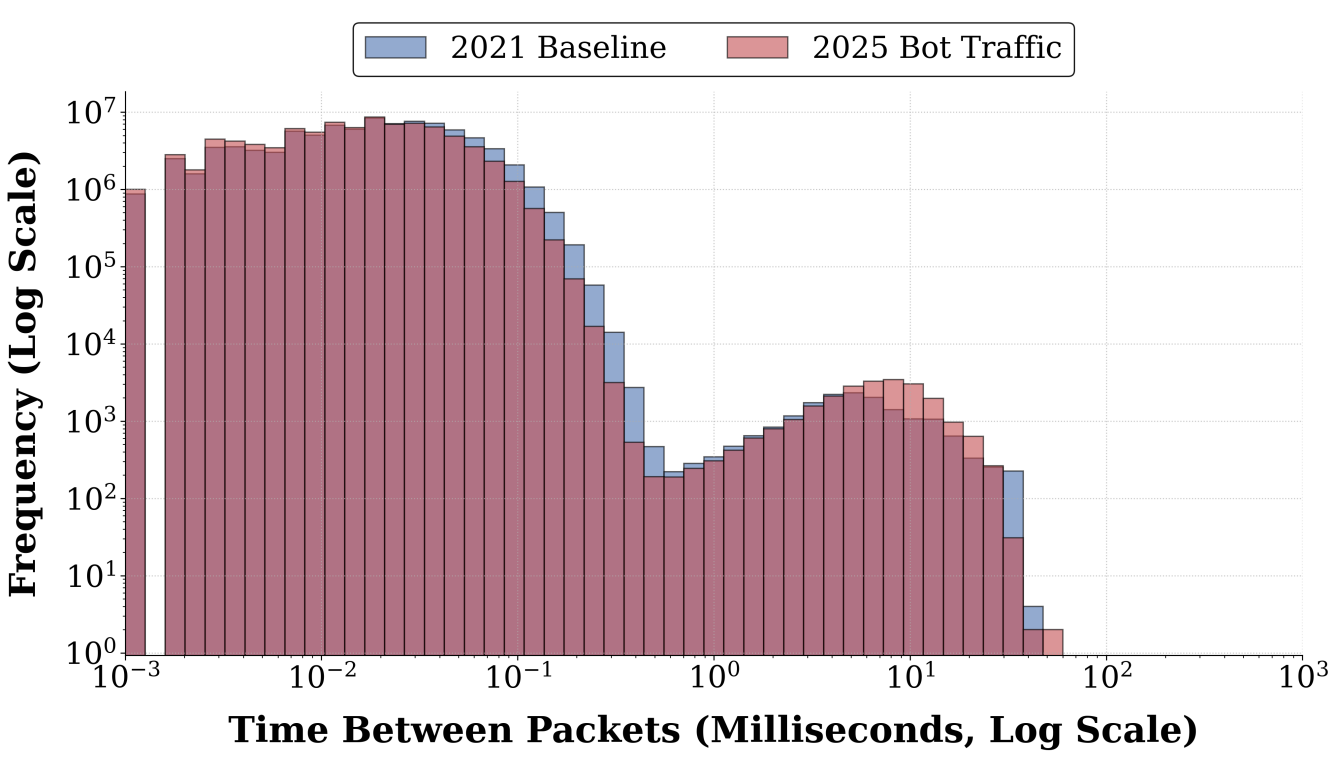

Fig. 4. Inter-Arrival Time (IAT) Distribution and Traffic Burstiness

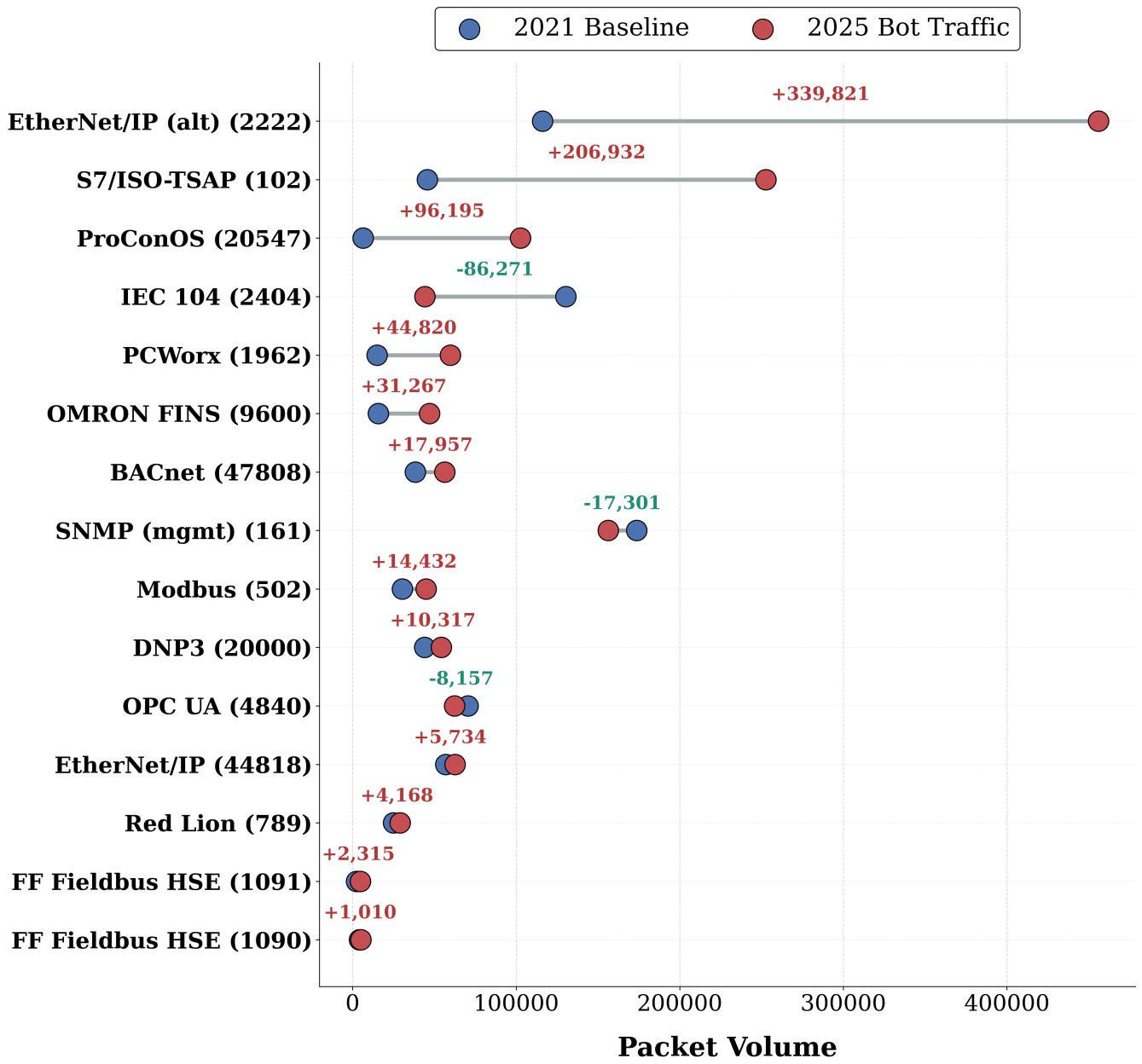

Fig. 5. Cross-Year Packet Volume Shifts Across Top Targeted Industrial Protocol Ports

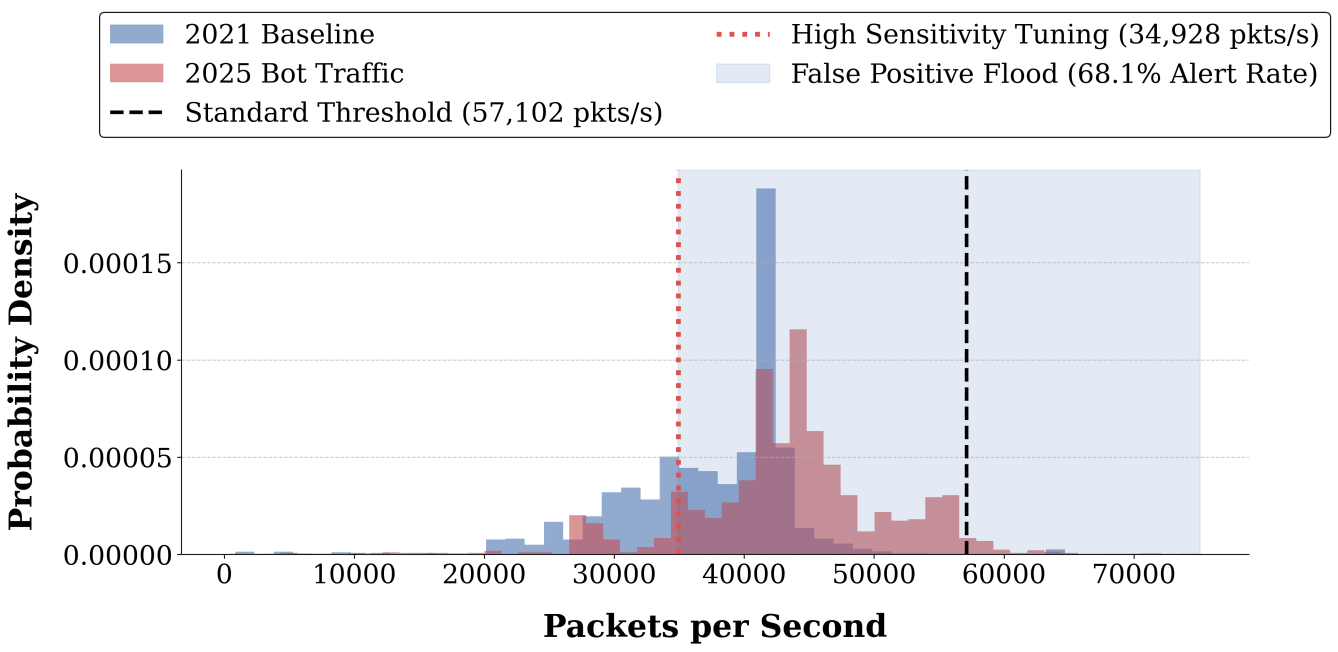

Fig. 6. IDS Anomaly Simulation: Volumetric Threshold and False Positives

- Explosive shifts in geographic origin, heavily utilizing localized regional infrastructures.
- Decreased destination port entropy indicating specific targeting of industrial protocols (resulting in a near-doubling of ICS traffic).
- Increased source IP entropy confirming a widely distributed attack swarm.
- Deliberate packet pacing (1ms to 100ms gaps) mapped via inter-arrival times.

These characteristics are consistent with increasing use of automated and potentially AI assisted scanning tools. From a security perspective, this structured behavior has important implications for intrusion detection systems. Specifically, the ability of modern botnets to smoothly blend their operations into historical network noise, evading over 97% of standard volumetric thresholds. This demands a shift toward behavior-aware and pacing-sensitive detection mechanisms for critical infrastructure.

## VI. Future Work

To build upon the conclusive findings of this longitudinal study, several extensions to this research are planned.

First, having demonstrated the failure of traditional volumetric anomaly detection (yielding a 97.47% evasion rate), our primary focus will shift toward developing next-generation, behavior-aware Intrusion Detection Systems (IDS) for OT environments. Future work will leverage the temporal burstiness and inter-arrival time (IAT) metrics identified in this study to train machine learning models [21]. These models will be designed to explicitly detect the artificial micro-pacing (1ms to 100ms delays) characteristic of modern botnets, thereby isolating intelligent reconnaissance without triggering the false-positive flood associated with static thresholds.

Second, while the Merit ORION Network Telescope provides visibility into passive background scanning, it inherently lacks visibility into post-connection exploitation. Future work will aim to correlate these massive passive telescope observations with data from geographically distributed, high-interaction ICS honeypots. This will allow us to capture and analyze the actual application-layer payloads being delivered to industrial ports (such as EtherNet/IP and S7/ISO-TSAP) once the initial automated reconnaissance phase is complete.

Third, we plan to transition the current offline, batch-processing analysis pipeline into a continuous, real-time streaming architecture. By integrating stream-processing frameworks, the core statistical evaluations, including dynamically scaling gap analysis, global Shannon entropy calculations, and pre-binned IAT tracking, can be executed live. This enhancement would enable the creation of a real-time threat intelligence dashboard and alerting feed specifically tailored for critical infrastructure operators.

Finally, further research is required to classify and potentially reverse-engineer the specific automated and AI-driven mechanisms dictating these distributed scanning campaigns. By applying advanced clustering algorithms to the regional geographic shifts and protocol-specific targeting patterns, we aim to map these distinct structural signatures to specific evolving botnet families or advanced persistent threat (APT) groups focusing on operational technology.

## VII. Conclusion

This work presented a reproducible, highly optimized pipeline that successfully ingested and analyzed a longitudinal dataset of 192 million passive darknet packets from the Merit ORION Network Telescope, spanning 2021 and 2025. By extracting packet features and computing metrics such as ICS/OT port usage, global entropy, IAT burstiness, and geographic distribution, we conclusively characterized the evolution of unsolicited automated traffic and its direct threat to critical infrastructure.

Our findings demonstrate a paradigm shift from broad, opportunistic internet scanning to structured, AI-assisted reconnaissance. Traffic targeting specific industrial protocols, most notably EtherNet/IP (alt) and S7/ISO-TSAP, nearly doubled over the four-year period. Furthermore, geographic attribution revealed a massive migration of scanning infrastructure toward regional clusters. Most critically, burstiness analysis isolated deliberate 1ms to 100ms micro-pacing behaviors utilized by these distributed botnets to artificially smooth their volumetric footprint.

When evaluated against a simulated volumetric anomaly-based intrusion detection system, this intentional pacing proved effective, allowing 97.47% of modern bot traffic to seamlessly evade standard detection baselines. Attempts to mitigate this evasion through high-sensitivity tuning resulted in an unsustainable 68.10% false-positive rate. Ultimately, this research proves that standard volumetric anomaly detection is fundamentally incapable of securing OT environments against contemporary automated threats, necessitating an industry shift toward behavior-aware and pacing-sensitive detection models.

## Acknowledgment

This research was partially funded by NSF under the CICI: TCR program, IRIS: Instrumentation for Research and Inter-institutional SOC, Award Number NSF 2319793. Additional support was partially provided by the Institute for Northwest Energy Futures (INEF) and Washington State University Tri-Cities.